\documentstyle[preprint,prb,aps]{revtex}
\begin{document}
\draft
\title{Electron Transport in Diborides: Observation of Superconductivity in
{\rm ZrB$_2$}\\}

\author{Vitaly A. Gasparov$^{+}$\/\thanks{e-mail: vgasparo@issp.ac.ru}, N.S.
Sidorov, I.I. Zver'kova, M.P. Kulakov }

\address { Institute of Solid State Physics RAS, 142432 Chernogolovka,
Moscow district, Russian Federation.\\}

\date{Submitted to JETP Letters, April 2, 2001}
\maketitle
\begin{abstract}
We report on syntheses and electron transport properties of polycrystalline
samples of diborides ({\rm AB$_2$}) with different transition metals
atoms ({\rm A=Zr,Nb,Ta}). The temperature dependence of resistivity,
$\rho(T)$, and {\it ac} susceptibility of these samples reveal
superconducting transition of {\rm ZrB$_2$} with {\rm T$_c=5.5~K$},
while {\rm NbB$_2$} and {\rm TaB$_2$}  have been observed nonsuperconducting
up to $0.37~K$. {\rm H$_{c2}$ (T)} is linear in temperature below {\rm
T$_c$}, leading to a rather low {\rm H$_{c2}(0)= 0.1~T$}. At {\rm T}  close
to {\rm T$_c$} {\rm H$_{c2}$ (T)} demonstrates a downward curvature. We
conclude that these diborides as well as {\rm MgB$_2$} samples behaves like a
simple metals in the normal state with usual Bloch-Gr\"uneisen temperature
dependence of resistivity and with Debye temperatures: $280~K$, $460~K$ and
$440~K$, for {\rm ZrB$_2$}, {\rm NbB$_2$} and {\rm MgB$_2$}, respectively,
rather than {\rm T$^2$} and {\rm T$^3$} as previously reported for {\rm
MgB$_2$}.
\end{abstract}


The recent discovery of Akimitsu \cite{nagamatsu01} of superconductivity in
{\rm MgB$_2$} at $39~K$ has lead to a booming activity in physics
community. It was reported that measurements of the upper critical field,
{\rm H$_{c2}$(T)}, and the thermodynamic critical field, {\rm H$_c$(T)},
as well as the specific heat are all consistent with {\rm MgB$_2$} being
a fairly typical intermetallic electron phonon mediated BCS superconductor
with a superconducting transition temperature approximately twice that of a
typical "metallic" superconductor. At the same time an unexpected {\rm
T$^2$}- and even {\rm T$^3$}- normal-state temperature dependence of the
resistance of has been reported \cite{jung01,finnemore01}.  These results
have initiated considerable interest to electron transport measurements and
to a search for superconductivity in other diborides.  Natural candidates for
this search are {\rm AB$_2$} - type light metals diborides ({\rm A =
Li,Be,Al}). However, up to now superconductivity has not been reported in the
majority of these compounds \cite{sluski01,felner01,zhao01}. Only very
recently has superconductivity below $1~K$ ($T_c=0.72~K$) been reported in
{\rm BeB$_{2.75}$} \cite{young01}. According to Ref. \cite{leyarovska79}, no
superconducting transition down to $0.42~K$ has been observed in powders of
diborides of transition metals ({\rm A = Ti,Zr,Hf,V,Ta,Cr, Mo,U}). Only {\rm
NbB$_2$} was found to be superconducting with very low {\rm T$_c = 0.62~K$}.
Since completing this work we have become aware of a recent paper
\cite{kaczorowski01} in which superconductivity was observed in {\rm TaB$_2$}
at {\rm T$_c = 9.5~K$}. This result contradict the data reported in this
letter.

The crystal structure of diborides ($\omega$ phase \cite{bianconi01}) can be
viewed as a set of two-dimensional graphite-like monolayers of {\rm Boron}
atoms with honeycomb lattice, intercalated with {\rm A}  atoms monolayers
forming {\rm ABAB}  heterostructure. It was proposed that superconductivity
in series of diborides of light metals ({\rm A=Be,Mg,Ca,Al}) and a series
transition metals ({\rm A=Ti,Zr,Hf,V,Nb,Ta}) appear near a critical point in
a strain and charge density phase diagram. An increase of the {\rm B-B}
distance larger than a critical value driven by {\rm A} atom radii results in
collapse of ideal $\omega$ phase into the trigonal rumpled $\omega$ phase.
Lattice instability driven by both atom radii and charge densities may be a key
reason why superconductivity appear in {\rm MgB$_2$} rather than in other
diborides.  This phase diagram suggests that {\rm ZrB$_2$} is a good
candidate for superconductivity having strain close to the critical value
(see \cite{bianconi01}). In order to clarify whether this assumption is
correct we studied the electron transport properties of {\rm ZrB$_2$,
NbB$_2$, TaB$_2$} and {\rm MgB$_2$}.

Polycrystalline samples of {\rm ZrB$_2$} and {\rm NbB$_2$} were obtained by
conventional solid state reaction. Starting materials were {\rm Zr} and {\rm
Nb} metal powders ($99.99\%$ purity) and sub-micron amorphous {\rm Boron}
powder ($99.9\%$ purity). These materials were lightly mixed in appropriate
amounts and pressed into a pellet $10~mm$ thick and $20~mm$ in diameter. The
pellets were placed on {\rm W} foil, which was in turn placed between two
{\rm W} ingots. These whole assembly was burned at $1700^{\circ}C$ for two
hours in an HV chamber at $2\times 10^{-4}~Pa$ by {\rm e}- beam heating of
{\rm W} ingot. The {\rm TaB$_2$} samples were prepared in the same way from
{\rm TaB$_2$} powder (Donezk Factory of Chemicals).

The resulting ceramic pellets had over $60 - 90\%$ the theoretical mass
density and were black in color. They demonstrated good metallic conductivity
at low temperatures. After regrinding the prepared pellets in an agate
mortar, the respective powders were placed in a tungsten can and melted by
{\rm e}-beam heating of the {\rm W} crucible for a few minutes in {\rm HV}
chamber at $1\times 10^{-4}~Pa$. Resulting melt had shiny single crystals
about $0.1~mm$ in size on top with solid polycrystalline ingot underneath
between melted {\rm W} can walls. However, these single crystals were found
to be nonsuperconducting due to presence of high concentration of {\rm W}
impurities observed from {\rm XRD}. The {\rm MgB$_2$} samples were sintered
from metallic {\rm Mg} plates placed on top of {\rm Boron}-pressed pellets.
The pellets were placed on {\rm Mo} foil, which was in turn placed on a {\rm
Mo} crucible. The crucible, with foil and pellets, was burned for two  hours
at $1400^{\circ}C$  in a tube furnace under an {\rm Ar} pressure of
$20~atm.$, with subsequent cooling to room temperature when furnace was
turned off.

X-ray $\theta -2\theta$ scan diffraction pattern (see Fig.\
\ref{fig1}) was obtained using {\rm CuK$_{\alpha}$} radiation. The
measurements showed the samples of all diborides to consist largely of the
desired {\rm AlB$_2$} phase. We would like to note that a small amount of
{\rm ZrB$_{12}$} impurity phase was found to be present after first step of
preparation. This phase was washed out after subsequent regrinding and
annealing. The {\rm X}-ray diffraction pattern was indexed within a hexagonal
unit cell (space group {\it P6/mmm}). This crystal structure consists of
honeycomb-net planes of {\rm Boron}, separated by triangular planes of the
metals. The analysis of the data yielded following values of the lattice
parameters: $a = 3.170~\AA$ and $c = 3.532~\AA$  for {\rm ZrB$_2$}, $a =
3.111~\AA$, $c =3.267~\AA$ for {\rm NbB$_2$}, and $a = 3.087~\AA$, $c =
3.247~\AA$ for {\rm TaB$_2$}. All parameters are in good agreement with
published data \cite{nbs83}.

For resistive measurements we used following sample preparation procedure:
the pellets were first cut by using a diamond saw with ethyl alcohol as
coolant. Rectangular solid bars of about $0.5 \times 0.5 \times 6~mm^3$ size
were obtained from pellets by using the spark erosion method. A standard
four-probe {\it ac} ($9~Hz$) method was used for resistance measurements.
Silver Print was used for making electrical contacts. The samples were mount
in a liquid helium cryostat able to regulate the temperature from $1.8~K$ to
room temperature. We employed a radio frequency ({\rm RF}) coil technique for
{\it ac} susceptibility measurement \cite{gasparov94}. Magnetoresistivity
measurements utilizing a superconducting magnet were also performed. A
well-defined geometry of the samples allowed us to perform accurate
measurements of resistivity with the magnetic field applied perpendicular to
the direction of electric current.

Fig.\ \ref{fig2} shows the typical temperature dependence of the resistivity
for {\rm ZrB$_2$} samples. The resistivity exhibits a very sharp
superconducting transition at $5.5~K$ shown on Fig.2b. The
transition width is about $0.14~K$ for a $10$ to $90\%$ drop. Such a narrow
transition  is a characteristic of good quality superconducting material. The
resistivity value $10~\mu\Omega \cdot cm$ at room temperature is almost the
same as that of {\rm MgB$_2$} (see Fig.\ \ref{fig4}), whereas the residual
resistivity is about $2~\mu\Omega\cdot cm$. One should note that  the
temperature dependence of the resistivity bellow $150~K$, may be fitted by a
square law, $\rho(T) = a+b \times T^2$, characteristic for electron-electron
scattering \cite{gasparov93}. Nevertheless, the overall temperature
dependence of resistivity can be perfectly fitted with classical
Bloch-Gr\"uneisen law (solid line), with rather low Debye temperature, $T_D =
280~K$.

Figure 2b  shows temperature dependence of the normalized real part of {\it
ac } susceptibility and the resistivity, $\rho(T)$, in the same {\rm ZrB$_2$}
sample as on Fig.2a. For this type of measurements the sample is
placed inside the inductance coil of {\rm LC} circuit, and the resonant
frequency of this circuit is monitored as a function of temperature. The
temperature dependence of the sample {\it ac } susceptibility $\chi(T)$
causes changes in inductance {\rm L} and in turn the resonant frequency
$f(T)$ of the circuit, $\Delta f/f\propto -Re\chi$.
Susceptibility data of Fig.2a clearly show the superconducting
transition in {\rm ZrB$_2$} sample at $5.5~K$. The onset temperature of the
{\it ac } susceptibility corresponds to the end point resistance transition,
typical for superconducting transition.

Fig.\ \ref{fig3} shows magnetic field dependent electrical resistivity data
taken at a variety of temperatures. Using these data, the resistive upper
critical magnetic field, {\rm H$_{c2}$(T)}, can be extracted from each
curve. We can do it by  extending the $\rho(H)$ line with maximum $d\rho/dH$
up to the normal state (see the dotted line). Figure 3b presents temperature
dependence of the {\rm H$_{c2}$(T)} for a {\rm ZrB$_2$} sample derived from
this extrapolation. Contrary to conventional theory \cite{helfand66}, we
found that the {\rm H$_{c2}$(T)} dependence is linear over an extended region
of {\rm T} with some downward curvature close to {\rm T$_c$}. This leads to a
rather low zero temperature value of {\rm H$_{c2}(0) =  0.1~T (\xi(0)=
570\AA$) which is significantly smaller than for {\rm MgB$_2$} ($18~T$)
\cite{budko01}. Such a linear {\rm H$_{c2}$(T)} dependence is characteristic
for 2D superconductors \cite{prober80} and was also observed in {\rm MgB$_2$}
\cite{budko01} wires and {\rm BaNbO$_{3-x}$} films
\cite{gasparov00,gasparov01}. The results of extended study of the {\rm
H$_{c2}$(T)} dependence {\rm ZrB$_2$} samples will be discussed elsewhere.

Fig.\ \ref{fig4} shows the temperature dependencies of the resistivity for
{\rm NbB$_2$} and {\rm MgB$_2$} samples. Best fits were obtained using
classical Bloch-Gr\"uneisen law rather than $a + b \times T^2$ or $a+b \times
T^3$ forms previously reported for {\rm MgB$_2$} \cite{jung01,finnemore01}.
The solid fitting lines with the Bloch-Gr\"uneisen form in Fig.\ \ref{fig4}
nearly overlays the data. The Debye temperatures derived from these fits are
rather high in comparison with {\rm ZrB$_2$}, i.e. $460~K$ and $440~K$ for
{\rm NbB$_2$} and {\rm MgB$_2$}, respectively. The $T_c$ value for the {\rm
MgB$_2$} samples is the same that is found elsewhere, i.e. $39~K$
\cite{nagamatsu01,jung01,finnemore01}. We suggest that previously reported
$T^2$  and $T^3$ dependencies of the resistivity of {\rm MgB$_2$} may be due
to rather high residual resistance, $\rho(0)$, of the samples and
respectively weak  $\rho(T)$ dependence. Our data are well consistent with
the Bloch-Gr\"uneisen law, which is characteristic for electron-phonon
scattering \cite{gasparov93}. Nevertheless, we believe that only  data
obtained on high purity single crystals can solidify this conclusion. Note
that the resistivity ratio for all diborides studied is approximately $5$
rather then $10000$ or even few millions as for pure metals
\cite{gasparov93}.

In contrast to the data of Ref. \cite{kaczorowski01} our {\it RF }
susceptibility measurements of {\rm TaB$_2$} samples did not reveal any
superconducting transition. At the same time our {\rm XRD} data have shown
the same \cite{kaczorowski01} clear  {\rm AlB$_2$} spectra (see Fig.\
\ref{fig1}).  In contrast to \cite{kaczorowski01}, we did not find any
impurities in our {\rm TaB$_2$} samples from XRD data (see Fig.\ \ref{fig1}).
Our resistive measurements up to $0.37~K$ do not show any superconducting
transition in {\rm NbB$_2$} as well. This contradicts the {\it ac }
susceptibility data of Ref. \cite{leyarovska79}, which reports the
superconducting transition in {\rm NbB$_2$}  to be at $0.62~K$. We observed
the superconducting transition at $4.4~K$ in melted {\rm TaB$_2$} samples
while the {\rm XRD} data of these samples revealed the presence of metallic
{\rm Ta} impurities. We believe these contradictions are due to the different
sample preparation technique of Refs. \cite{leyarovska79}and
\cite{kaczorowski01}. The samples used in those papers were  prepared using
a borothermic method from {\rm Nb$_2$O$_5$}  and {Ta$_2$O$_5$} pentoxides and
{\rm Boron} \cite{peshev68}. Therefore, those samples might consist of
reduced {\rm Nb} and {\rm Ta} oxides intercalated by {\rm Boron}, which could
be the source of superconducting transition. Recently, we have shown that
oxygen reduced {\rm BaNbO$_{3-x}$} compounds exhibit superconducting
transition even at $T_c=22~K$ \cite{gasparov00,gasparov01}. Also, since the
superconductivity in {\rm TaB$_2$} was not observed before on the samples
made by the same technique \cite{leyarovska79},  the reason of this
discrepancy with data obtained in \cite{kaczorowski01} should be emphasized
from resistance measurements.  Therefore, care must be taken for sample
preparation and composition before final conclusions about superconductivity
in {\rm NbB$_2$}, {\rm TaB$_2$} as well in other diborides can be drawn.

{\it In summary}, we report the temperature dependence resistivity and high
frequency susceptibility measurements of synthesized samples of {\rm
ZrB$_2$,NbB$_2$,TaB$_2$} and {\rm MgB$_2$}. We discovered a superconducting
transition at $5.5~K$ in {\rm ZrB$_2$} samples. The superconducting
transition width for the resistivity measurement was about $0.14~K$, and the
resistivity in the normal state followed a classical Bloch-Gr\"uneisen
behavior in all diborides studied. The linear {\rm H$_{c2}$(T)} dependence
was observed for {\rm ZrB$_2$} samples over an almost full range of
temperatures below {\rm T$_c$}, leading to a rather low $H_{c2}(0) = 0.1~T$.
Superconducting transition was not found in {\rm NbB$_2$} samples up to
$0.37~K$ and {\rm TaB$_2$} samples above $4.4~K$.

\acknowledgments

We gratefully acknowledge helpful discussions with V.F. Gantmakher,
measurements of $\rho(T)$ dependence in {\rm NbB$_2$} samples up to $0.37~K$
with V.N. Zverev, help in manuscript preparation with L.V. Gasparov, help in
samples preparations with S.N. Ermolov, V.V. Lomejko and V.G. Kolishev. This
work was supported by the Russian Council on High-Temperature
Superconductivity (Grant No. Volna 4G) and the Russian Scientific Program:
Surface Atomic Structures (Grant No.4.10.99).

\newpage

\begin{figure}
\caption{ {\rm X}-ray $\theta-2\theta$ scan of {\rm ZrB$_2$}, {\rm NbB$_2$},
and {\rm TaB$_2$} pellets,  at room temperature. The cycles mark the
reflections from cubic {\rm ZrB$_{12}$} ($a = 7.388~\AA$) impurity.  }
\label{fig1}
\end{figure}

\begin{figure}
\caption{Fig.2a. Typical temperature dependence of the resistivity of {\rm
ZrB$_2$} sample. Symbols denote the data and the solid line is
Bloch-Gr\"uneisen law fit with Debye temperature $280~K$. Fig.2b.
Low-temperature dependence of the {\it ac} susceptibility (proportional to
the frequency shift [$f(T)-f_0]/f_0$- left axis) measured with the  frequency
$f_0  =  f(6K) = 9~MHz$, and the resistivity (right axis) in ZrB$_2$.}
\label{fig2}
\end{figure}

\begin{figure}
\caption{ Fig.3a. Magnetic field resistive transition curves in {\rm ZrB$_2$}
sample at different temperatures. The dashed line is drawn through the
$\rho(H)$  dependence with maximal $d\rho/dH$ derivative for illustration of
the {\rm H$_{c2}$} determination. Fig.3b. Temperature variation of the upper
resistive critical magnetic field {\rm H$_{c2}$(T)} in {\rm ZrB$_2$} sample.}
\label{fig3}
\end{figure}

\begin{figure}
\caption{Typical temperature dependence of the resistivity of {\rm NbB$_2$}
and {\rm MgB$_2$} samples. The solid lines denotes the Bloch-Gr\"uneisen law
fit with Debye temperatures $440~K$ and $460~K$ for {\rm MgB$_2$} and {\rm
NbB$_2$}, respectively. }
\label{fig4}
\end{figure}

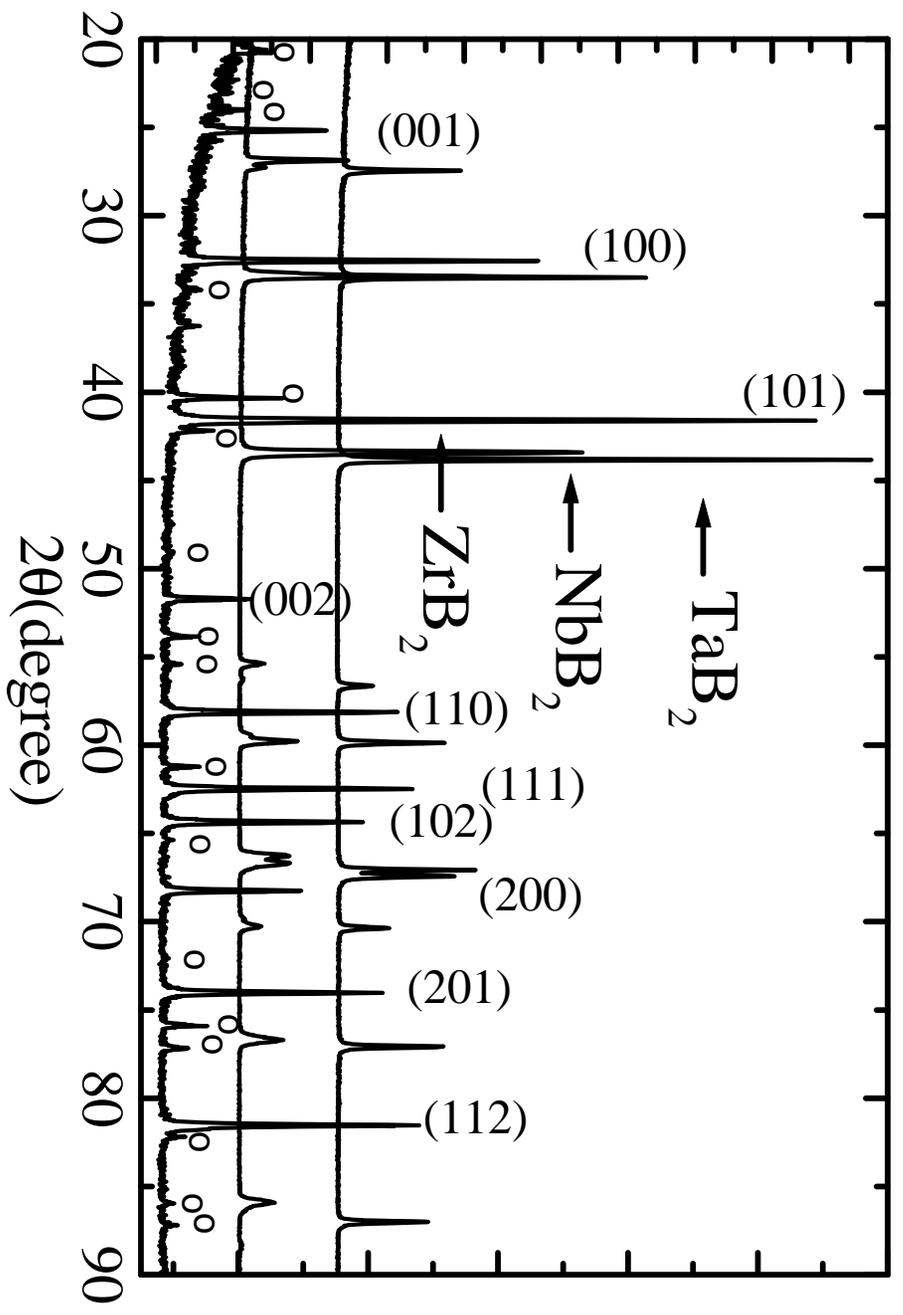
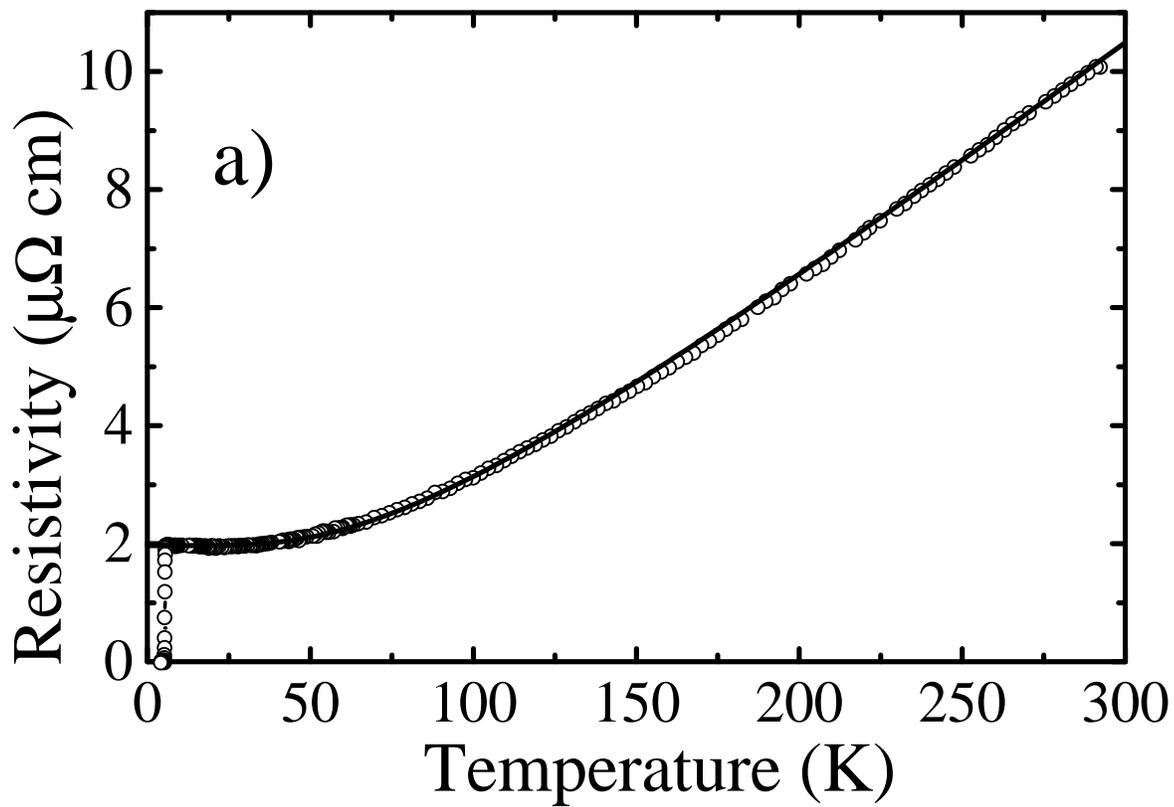
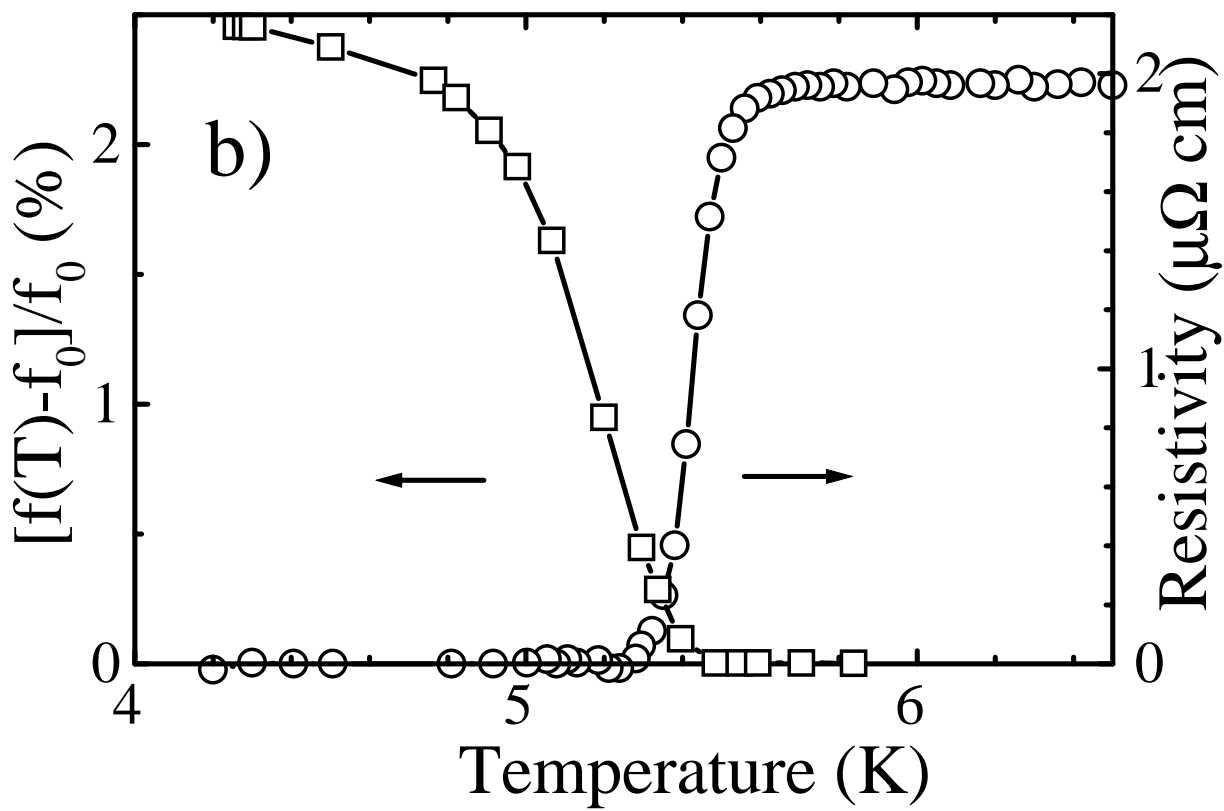
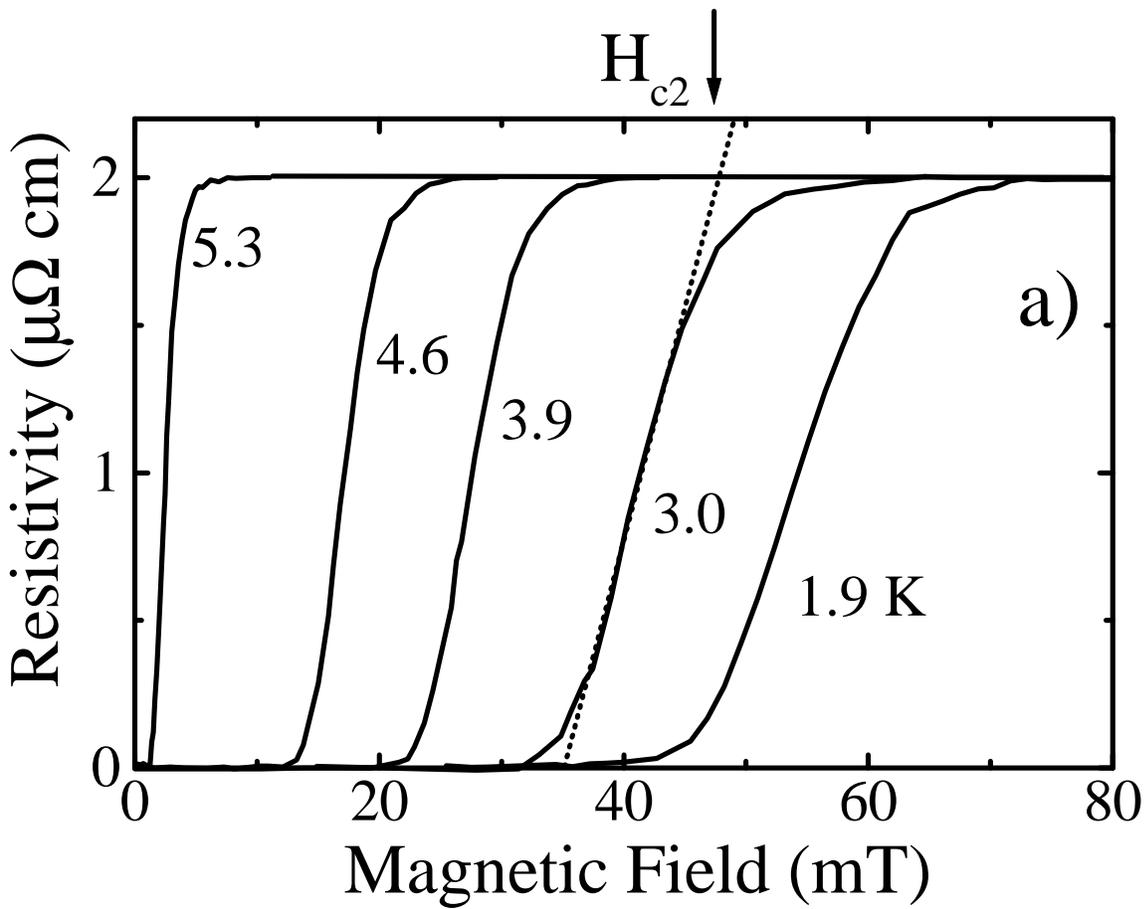
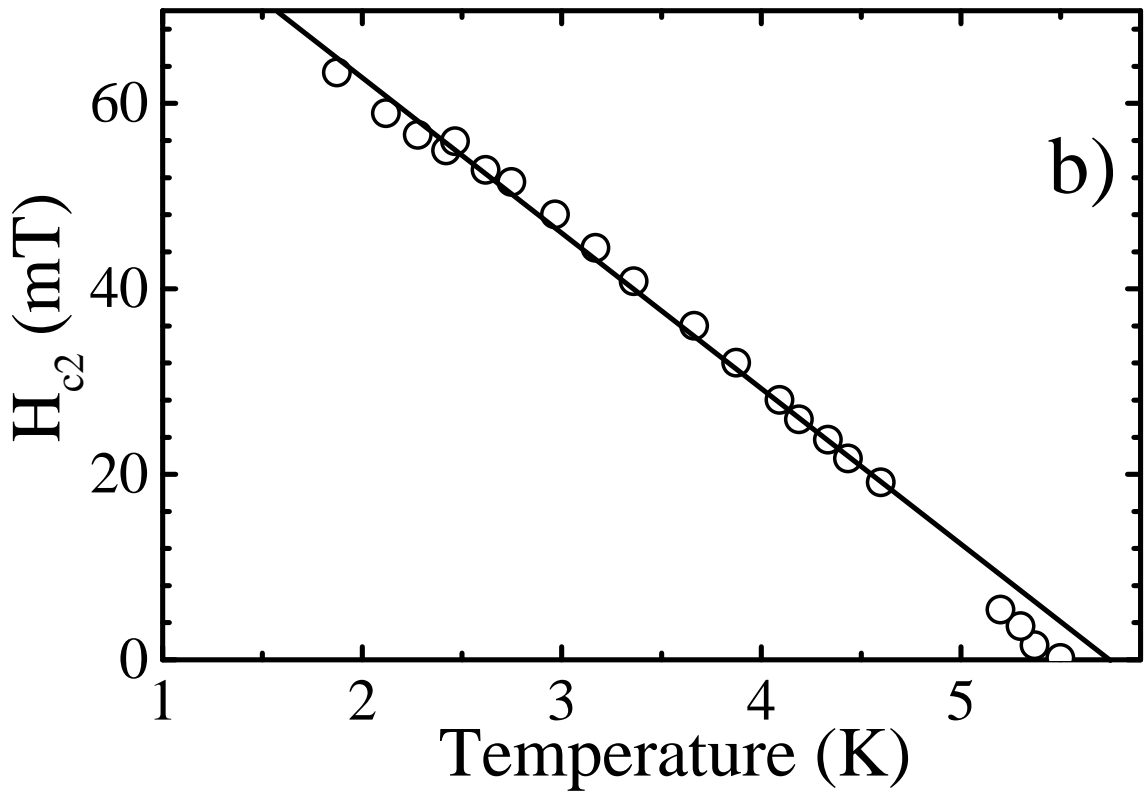
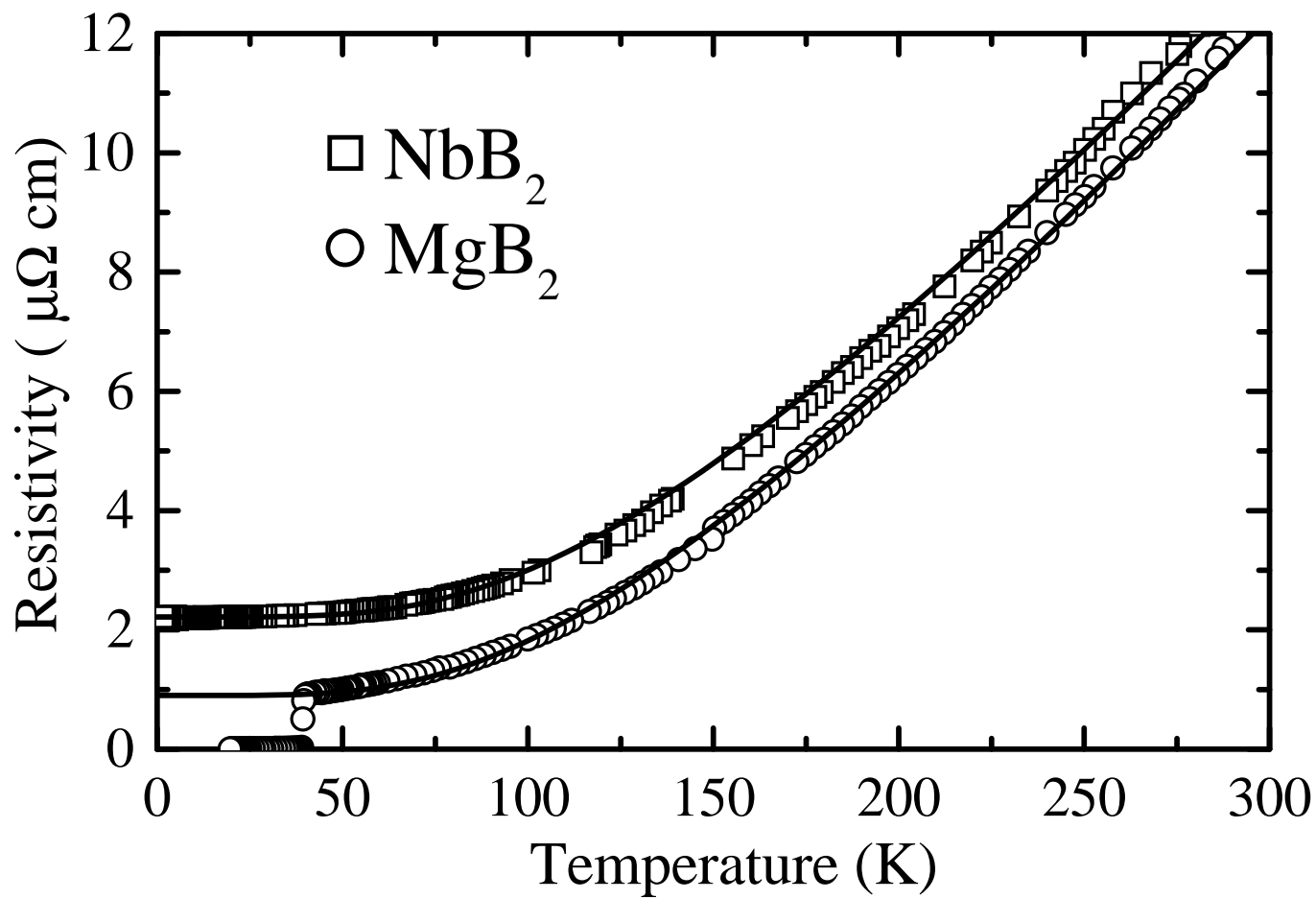
\end{document}